\documentclass[twocolumn,aps,pra,superscriptaddress]{revtex4-1}
\usepackage[utf8]{inputenc}
\usepackage{amsmath}
\usepackage{amssymb}
\usepackage{bm}
\usepackage{graphicx}
\usepackage{color}

\newcommand{\Tr}{\mathop{\mathrm{Tr}}\nolimits}
\newcommand{\grad}{\boldsymbol{\nabla}}

\begin{document}

\title{Superconductor-insulator transition in Josephson junction chains by quantum Monte-Carlo}

\author{D. M. Basko}

\affiliation{Laboratoire de Physique et Mod\'elisation des Milieux Condens\'es, Universit\'e Grenoble Alpes and CNRS, 25 rue des Martyrs, 38042 Grenoble, France}

\author{F. Pfeiffer}
\altaffiliation[Present address: ]{Physics Department, Arnold Sommerfeld Center for Theoretical Physics, Ludwig-Maximilians-Universit\"at M\"unchen, 80333 M\"unchen, Germany}

\affiliation{Fachbereich Physik, Universität Konstanz, D-78457 Konstanz, Germany}
\affiliation{Laboratoire de Physique et Mod\'elisation des Milieux Condens\'es, Universit\'e Grenoble Alpes and CNRS, 25 rue des Martyrs, 38042 Grenoble, France}

\author{P. Adamus}

\affiliation{Department of Condensed Matter Physics, Faculty of Science, and Central European Institute of Technology, Masaryk University, Kotl\'a\v{r}sk\'a 2, 611 37 Brno, Czech Republic}
%\affiliation{Laboratoire de Physique et Mod\'elisation des Milieux Condens\'es, Universit\'e Grenoble Alpes and CNRS, 25 rue des Martyrs, 38042 Grenoble, France}

\author{M. Holzmann}

\affiliation{Laboratoire de Physique et Mod\'elisation des Milieux Condens\'es, Universit\'e Grenoble Alpes and CNRS, 25 rue des Martyrs, 38042 Grenoble, France}

\author{F. W. J. Hekking}
\altaffiliation[]{Deceased 15 May 2017}

\affiliation{Laboratoire de Physique et Mod\'elisation des Milieux Condens\'es, Universit\'e Grenoble Alpes and CNRS, 25 rue des Martyrs, 38042 Grenoble, France}

\begin{abstract}
We study the zero-temperature phase diagram of a dissipationless and disorder-free Josephson junction chain. Namely, we determine the critical Josephson energy below which the chain becomes insulating, as a function of the ratio of two capacitances: the capacitance of each Josephson junction and the capacitance between each superconducting island and the ground. We develop an imaginary-time path integral Quantum Monte-Carlo algorithm in the charge representation, which enables us to efficiently handle the electrostatic part of the chain Hamiltonian. We find that a large part of the phase diagram is determined by anharmonic corrections which are not captured by the standard Kosterlitz-Thouless renormalization group description of the transition.
\end{abstract}

\maketitle

\section{Introduction}

Josephson junction (JJ) chains are essential elements of many superconducting circuits, where microwave signals can propagate with little or no dissipation~\cite{Jung2014}. They are interesting both for applications, such as
metrological current standard~\cite{Bylander2005},
qubit protection from charge noise~\cite{Manucharyan2009},
building high-impedance environments \cite{Corlevi2006, Masluk2012, Bell2012},
or parametric microwave amplification \cite{Castellanos2008, Macklin2015, Planat2019}, as well as for studying fundamental phenomena, such as 
macroscopic quantum tunnelling \cite{Pop2010, Manucharyan2012, Ergul2013, Ergul2017},
phase-charge duality~\cite{Corlevi2006},
or strong-coupling quantum electrodynamics~\cite{Puertas2019}.

At the same time, JJ chains have been predicted to undergo a transition into an insulating state if the Coulomb energy associated with the transfer of a single Cooper pair is sufficiently high~\cite{Bradley1984, Korshunov1989}. Subsequently, such transition was observed experimentally \cite{Chow1998, Haviland2000, Haviland2001, Kuo2001, Myazaki2002, Takahide2006}. Random pinning of the insulator by disorder was suggested to be a fundamental obstacle~\cite{Cedergren2017} to realization of a metrological current standard based on quantum phase slip junctions~\cite{Mooij2006,Guichard2010}.

Given the importance of the problem and the high degree of control achieved in JJ chain fabrication, precise information about the insulating region in the parameter space would be highly desirable. Surprisingly, a quantitative theoretical prediction for the phase diagram is still lacking. Mappings between the quantum JJ chain, the classical two-dimensional (2D) $XY$ model, the 2D Coulomb gas, and the sine-Gordon model \cite{Bradley1984, Korshunov1989, Bobbert1990, Bobbert1992, Hermon1996, Choi1998, Gurarie2004, Ribeiro2014, Andersson2015, Bard2017} yielding an effective description of the system at long distances and low energies, established that the transition belongs to the Kosterlitz-Thouless universality class~\cite{Kosterlitz1973, Kosterlitz1974}. However, to precisely relate the parameters of an effective theory (e.~g., the Coulomb gas fugacity) to those of the physical JJ chain, one has to properly account for all short-distance contributions. This is possible only in some limiting cases.

\begin{figure}
\includegraphics[width=0.48\textwidth]{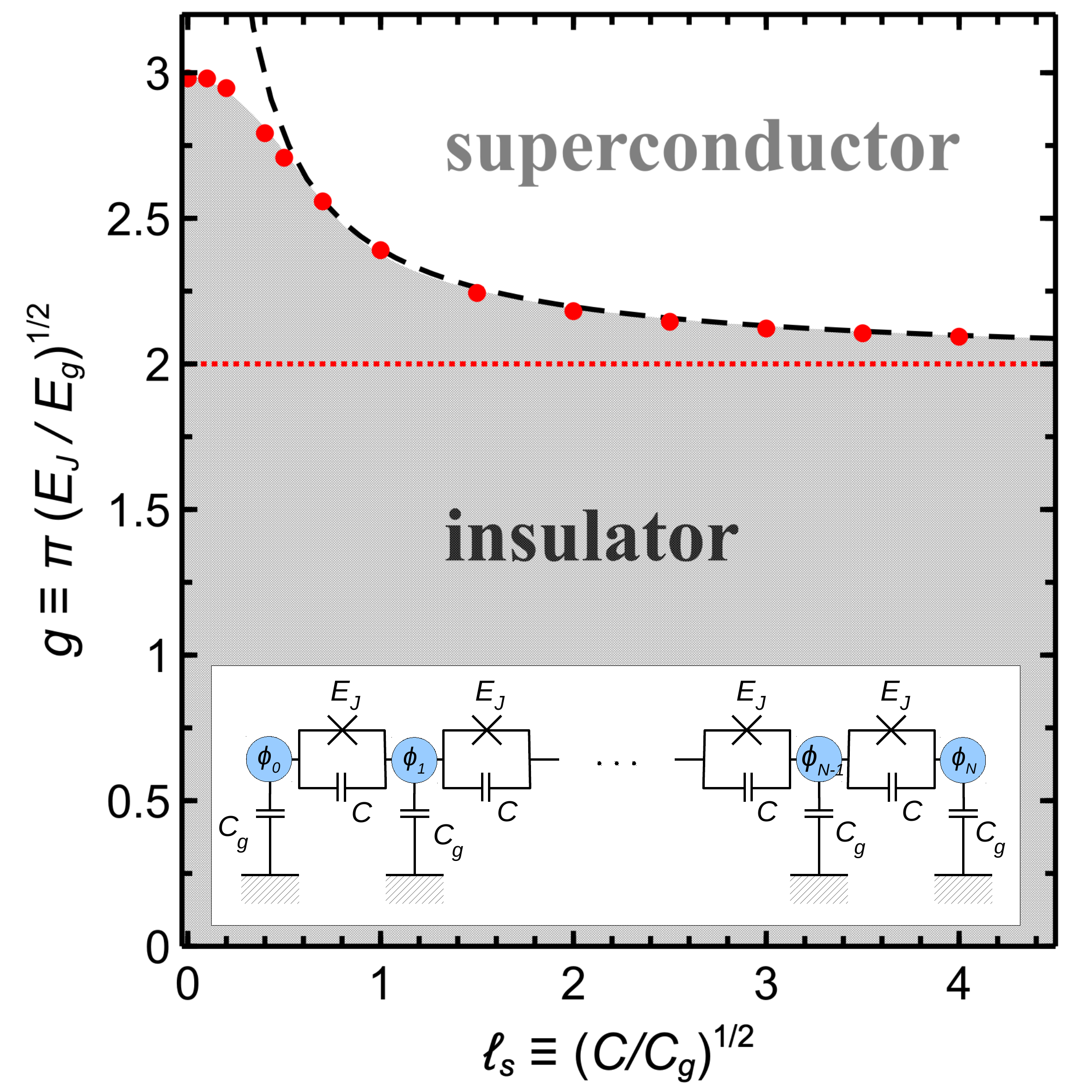}
\caption{The zero temperature phase diagram in the variables $\ell_s=\sqrt{C/C_g}$ and $g=\pi\sqrt{E_J/E_g}$, where $E_g\equiv(2e)^2/C_g$. The dots show our QMC results (the error bars are smaller than the symbol size). The horizontal dotted line $g=2$ is the $\ell_s\to\infty$ asymptote~\cite{Korshunov1989}, the dashed line corresponds to $g=2+\pi/(8\ell_s)$, discussed in Sec.~\ref{sec:asymptote}.
The inset shows schematically the JJ chain, described by Hamiltonian~(\ref{eq:H=}).}
\label{fig:JJchain}
\end{figure}

In the present paper, we start to fill this gap in the theoretical knowledge and numerically calculate the zero-temperature phase diagram of an isolated JJ chain in absence of any disorder and dissipation. We adopt the standard description of such chains as a long array of identical superconducting islands with Josephson and capacitive coupling between neighboring islands (characterized by the Josephson energy~$E_J$ and capacitance~$C$) and capacitive coupling between each island and a nearby ground plane (capacitance~$C_g$), as schematically shown in Fig.~\ref{fig:JJchain} (inset). The critical value of~$E_J$ was previously known only for $C/C_g\to\infty$~\cite{Korshunov1989, Choi1998, Rastelli2013} and numerical results were available for $C=0$ \cite{Danshita2011,Roscilde2016,Pino2016}; here we calculate it for an arbitrary ratio $C/C_g$ (Fig.~\ref{fig:JJchain}). We find that for $C/C_g\gtrsim1$ the critical $E_J$ is determined by the weak Kerr nonlinearity of the Josephson coupling~\cite{Weissl2015,Krupko2018}, a short-distance effect not captured by the standard Kosterlitz-Thouless renormalization group (RG) approach to the transition\cite{Kosterlitz1974, Bradley1984, Korshunov1989, Gurarie2004, Bard2017}.

To detect the transition, we develop a novel quantum Monte-Carlo (QMC) algorithm which evaluates directly the imaginary-time path integral in the charge representation, in contrast to the phase representation~\cite{Roscilde2016, Garanin2016} or Coulomb gas representation~\cite{Bobbert1992, Andersson2015}, used in previous works.
Our QMC scheme efficiently treats the Coulomb interaction and can be easily extended to include more complex electrostatic coupling~\cite{Krupko2018} or random offset charges~\cite{Ivanov2001, Gurarie2004, Vogt2015, Vogt2016, Bard2017}.

\section{The QMC scheme}

\subsection{Hamiltonian}

We consider a linear JJ chain consisting of $N+1$ identical superconducting islands labeled by an integer $n=0,1,\ldots,N$. The superconducting phase $\hat\phi_n$ of island~$n$ and the charge $\hat{q}_n$ on the island are canonically conjugate and satisfy the commutation relation $[\hat{q}_n,\hat\phi_{n'}]=2ie\delta_{nn'}$ ($e<0$ being the electron charge). We assume the chain to be fully isolated from the outside world, so the phases are compact and the charges are discrete. The chain is described by the Hamiltonian
\begin{equation}\label{eq:H=}
\hat{H}=\sum_{n,n'=0}^N\frac{C_{nn'}^{-1}}2\,\hat{q}_n\hat{q}_{n'}
+\sum_{n=1}^NE_J[1-\cos(\hat\phi_n-\hat\phi_{n-1})].
\end{equation}
While the last term represents the Josephson coupling between neighboring islands characterized by the Josephson energy~$E_J$, the first term describes the Coulomb interaction between the island charges. $C_{nn'}^{-1}$ is the inverse of the capacitance matrix; the latter is taken to be tridiagonal. The main diagonal is given by $C_{00}=C_{NN}=C_g+C$ and $C_{nn}=C_g+2C$ for $n=1,\ldots,N-1$, while the first diagonals are $C_{n,n-1}=C_{n-1,n}=-C$ for $n=1,\ldots,N$. Here $C_g$ and $C$ are the capacitances between each island and the ground, and between neighboring islands, respectively (Fig.~\ref{fig:JJchain}, inset). 
For $n,n'$ sufficiently far from the chain ends, the interaction falls off exponentially:
\begin{equation}\label{eq:Cnn-1=}
C_{nn'}^{-1}\approx\frac{1}{\sqrt{4CC_g+C_g^2}}
\left(1+\frac{C_g}{2C}-\sqrt{\frac{C_g}C+\frac{C_g^2}{4C^2}}\right)^{|n-n'|}.
\end{equation}
At $C=0$, the interaction is strictly local ($C_{nn'}^{-1}$ is proportional to the unit matrix).
For $C\gg{C}_g$, Eq.~(\ref{eq:Cnn-1=}) becomes
\begin{equation}
C_{nn'}^{-1}\approx\frac{{e}^{-|n-n'|/\ell_s}}{\sqrt{4CC_g}},
\quad\ell_s\equiv\sqrt{C/C_g}\gg{1},
\end{equation}
the screening length~$\ell_s$ determining the interaction range.

It is convenient to pass from the phases $\hat\phi_0,\ldots,\hat\phi_N$ defined on islands to phase differences defined on junctions, labeled by half-integers $j=1/2,3/2,\ldots,N-1/2$:
\begin{equation}
\hat\theta_{1/2}=\hat\phi_1-\hat\phi_0,\;
%\hat\theta_{3/2}=\hat\phi_2-\hat\phi_1,\ldots,\;
\hat\theta_{N-1/2}=\hat\phi_N-\hat\phi_{N-1},\;\hat\Phi=\hat\phi_N,
\end{equation}
$\hat\Phi$ being the global phase.
The corresponding conjugate variables, $\hat{P}_{1/2},\ldots,\hat{P}_{N-1/2},\hat{Q}$, are
\begin{equation}
\hat{P}_j=-\sum_{n<j}\hat{q}_n,\quad
\hat{Q}=\sum_{n=0}^N\hat{q}_n.
\end{equation}
$\hat{P}_j$ are the lattice analogs of the dielectric polarization field $\mathbf{P}$ (since $\hat{q}_n=\hat{P}_{n-1/2}-\hat{P}_{n+1/2}$, analogous to the charge density in a continuous medium, $\rho=-\grad\cdot\mathbf{P}$), while $\hat{Q}$ is the total charge of the chain.

In the following we focus on the sector $Q=0$, assuming the chain to be overall neutral. This assumption deserves some discussion.
The operators $\hat{q}_n$ represent the charge of the Cooper pair condensate, relative to the background charge of the grain, so they have positive and negative eigenvalues, integer multiples of $2e$, since each island can host an integer number of Cooper pairs. Here we assumed that the background charge of each grain is also an integer multiple of $2e$. This assumption can be relaxed by adding a term $-\sum_nV_{n}^g\hat{q}_n$, where the gate voltages $V_n^g$ can be the same for all islands or random. This gives rise to a rich variety of possible phases, whose study is beyond the scope of the present paper.
Restricted to the $Q=0$ sector, Hamiltonian~(\ref{eq:H=}) can be written as
\begin{equation}\label{eq:Htheta=}
\hat{H}_{Q=0}=\sum_{j,j'=1/2}^{N-1/2}\frac{D_{jj'}}2\,\hat{P}_j\hat{P}_{j'}
+E_J\sum_{j=1/2}^{N-1/2}(1-\cos\hat\theta_j),
\end{equation}
where
\begin{equation}\label{eq:Djj=}
D_{jj'}=\sum_{\sigma,\sigma'=\pm1}\sigma\sigma'C^{-1}_{j+\sigma/2,j'+\sigma'/2}
\end{equation}
is the dipole-dipole interaction matrix.

\subsection{Path integral}

To construct the imaginary-time path integral, we follow the standard procedure. Introducing a finite temperature $1/\beta$ (eventually to be extrapolated to zero) and splitting the imaginary time interval $0\leqslant\tau<\beta$ into $M\gg{1}$ slices of length $\varepsilon\equiv\beta/M$, we write the partition function as
\begin{equation}
\Tr\left\{e^{-\beta\hat{H}_{Q=0}}\right\}=
\Tr\left\{e^{-\varepsilon\hat{H}_{Q=0}}\ldots\,e^{-\varepsilon\hat{H}_{Q=0}}\right\},
\end{equation}
at each slice insert the unit operator in the $Q=0$ sector,
\[
\prod_{j=1/2}^{N-1/2}\sum_{P_j=-\infty}^\infty|P_j\rangle\langle{P}_j|,
\]
and approximate $e^{-\varepsilon\hat{H}_{Q=0}}=e^{-\varepsilon\hat{H}_C/2}e^{-\varepsilon\hat{H}_J}e^{-\varepsilon\hat{H}_C/2}+O(\varepsilon^3)$, where $\hat{H}_C$ and $\hat{H}_J$ are the Coulomb and the Josephson terms of the Hamiltonian in Eq.~(\ref{eq:Htheta=}), in order to evaluate the matrix element between different $|P_j\rangle$, the eigenstate of $P_j$. The Coulomb Hamiltonian $\hat{H}_C$ is diagonal in the $P_j$~basis, so $e^{-\varepsilon\hat{H}_C/2}$ gives just a numerical factor.
The matrix element of $e^{-\varepsilon\hat{H}_J}$ splits into a product over all junctions, each one contributing a factor
\begin{align}
\langle{P}|e^{\varepsilon{E}_J\cos\hat\theta}|P'\rangle={}&{}
\int_0^{2\pi}\frac{d\theta}{2\pi}\,
e^{i(l-l')\theta+\varepsilon{E}_J\cos\theta}\nonumber\\
={}&{}I_{l-l'}(\varepsilon{E}_J),\quad l\equiv\frac{P}{2e},\quad 
l'\equiv\frac{P'}{2e},
\end{align}
where $I_{l-l'}(z)$ is the modified Bessel function; note that $l,l'$ are integers.
We do not make the Villain approximation, $e^{\varepsilon{E}_J(\cos\theta-1)}\to\sum_m e^{-(\varepsilon{E}_J/2)(\theta-2\pi{m})^2}$ \cite{Villain1975, Janke1986}, often used to simplify the Josephson term \cite{Zwerger1989, Bobbert1990, Bobbert1992, Rastelli2013}. Working directly with Bessel functions, although formally beyond the $O(\varepsilon^3)$ precision, eliminates at least one source of errors at essentially no computational cost: since $I_l(\varepsilon{E}_J)$ quickly decreases with $l$ for $\varepsilon{E}_J\lesssim1$, only a few first orders $l$ of $I_l(\varepsilon{E}_J)$ are needed; they are calculated and stored before each QMC run.

\begin{figure}
\includegraphics[width=0.48\textwidth]{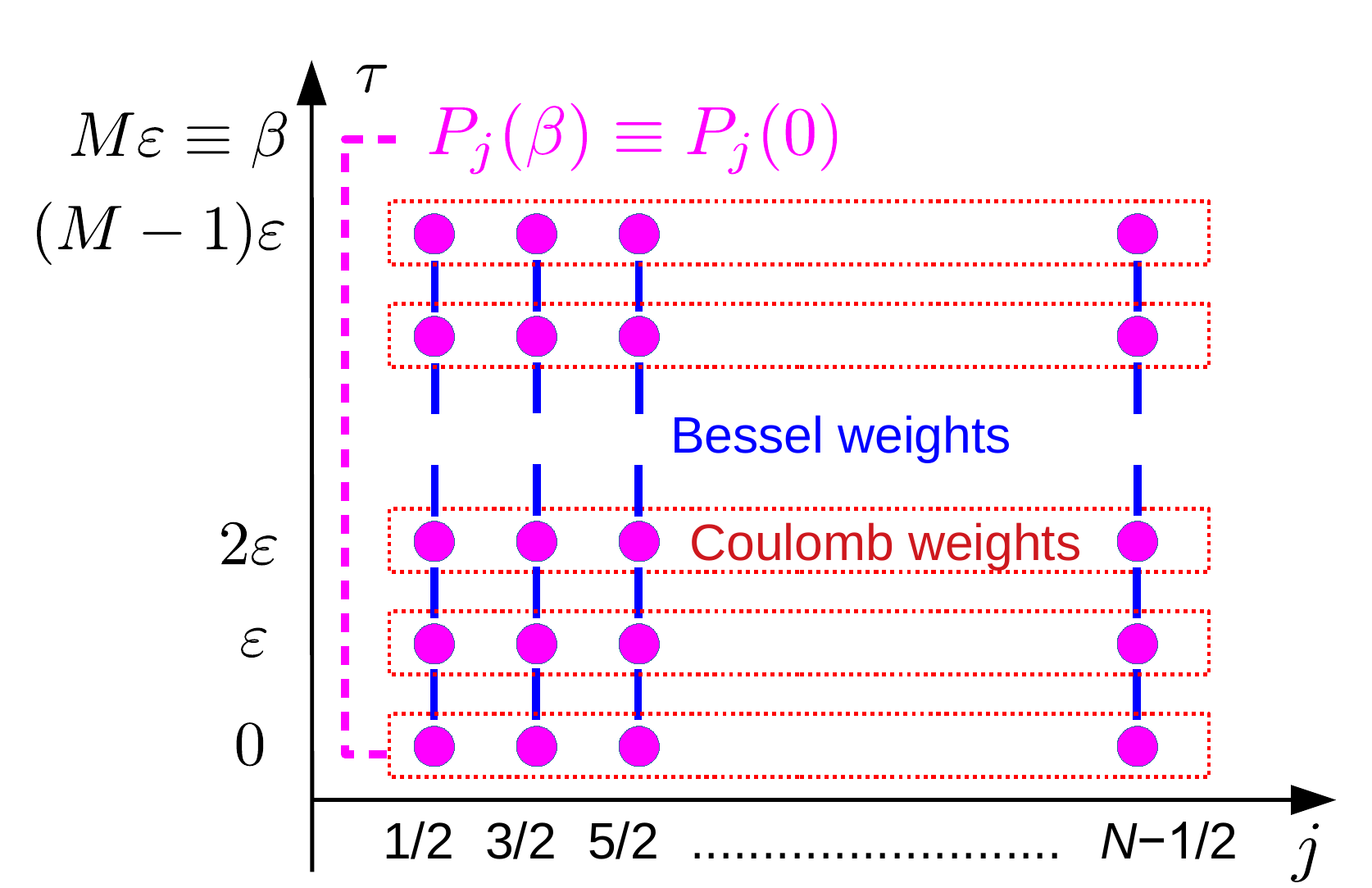}
\caption{A pictorial representation of Eqs.~(\ref{eq:QMC}). Each pink circle represents an integer summation variable $l_{jm}$ with periodic boundary conditions in the imaginary time, $l_{j0}\equiv{l}_{jM}$. Each blue segment corresponds to the the Bessel function $I_{l_{jm}-l_{j,m+1}}(\beta{E}_J/M)$. Each red dotted box corresponds to a Coulomb dipole sum at a given~$m$.}
\label{fig:QMC}
\end{figure}

As a result, for each given~$M$, the approximate partition function can be written as an $NM$-fold sum over integer variables $l_{jm}$:
\begin{subequations}\label{eq:QMC}\begin{align}
&\Tr\left\{e^{-\beta\hat{H}_{Q=0}}\right\}=e^{-\beta{N}E_J}
\lim_{M\to\infty}\sum_{\{l_{jm}\}=-\infty}^\infty W_CW_J,
\label{eq:QMCsum=}\\
&W_C=\exp\left(-\frac{(2e)^2\beta}{2M}\sum_{m=0}^{M-1}\sum_{j,j'=1/2}^{N-1/2}
D_{jj'}l_{jm}l_{j'm}\right),\\
&W_J=\prod_{j=1/2}^{N-1/2}\prod_{m=0}^{M-1}I_{l_{jm}-l_{j,m+1}}(\beta{E}_J/M),
\end{align}\end{subequations}
where we defined $l_{jM}\equiv{l}_{j0}$, so that the configurations are effectively on a cylinder. This construction is schematically represented in Fig.~\ref{fig:QMC}. It is straightforward to represent imaginary-time correlators of $\hat{P}_j$ operators in a similar way; for example,
$\Tr\{e^{-\beta\hat{H}_{Q=0}}e^{\tau\hat{H}_{Q=0}}\hat{P}_je^{-\tau\hat{H}_{Q=0}}e^{\tau'\hat{H}_{Q=0}}\hat{P}_{j'}\,e^{-\tau'\hat{H}_{Q=0}}\}$
%e^{-\beta{N}E_J}\lim_{M\to\infty}\sum_{\{l_{jm}\}=-\infty}^\infty W_CW_J(2e)^2l_{jm}l_{j'm'}
is given by the same sum~(\ref{eq:QMCsum=}), but with the summand $W_CW_J(2e)^2l_{jm}l_{j'm'}$ where $m,m'$ are such that $m\beta/M$ and $m'\beta/M$ are close to $\tau$ and $\tau'$, respectively. Correlators of $e^{\pm{i}\hat\theta_j}$ can also be calculated by inserting extra time slices and evaluating the corresponding matrix elements between the eigenstates of $\hat{P}_j$.

The $NM$-fold sum over the configurations $\{l_{jm}\}$ is evaluated by Monte-Carlo sampling of $W_CW_J$ with the standard Metropolis algorithm. To update the configuration, we use the following rule. First, we choose at random a junction $j$ and a segment $m_1\leqslant{m}\leqslant{m}_2$ on the imaginary-time circle (that is, one may have $m_2<m_1$, in which case the concerned variables are $l_{j,m\leqslant{m}_2}$ and $l_{j,m\geqslant{m}_1}$). The proposed new configuration is obtained by shifting $l_{jm}\to{l}_{jm}+\sigma$ on the chosen interval, with $\sigma=\pm1$ chosen randomly but the same for all~$m$ in the interval. This update modifies only two Bessel functions constituting the weight~$W_J$; since $I_{l-l'}(\varepsilon{E}_J)\sim (\varepsilon{E}_J)^{|l-l'|}$ at small $\varepsilon{E}_J$, an update modifying many Bessel functions would be likely to produce many small factors resulting in very low acceptance probability. Our rule results in the acceptance ratio of a few percent.  The change in the weight $W_C$ is calculated straightforwardly; it represents the main computational cost. For largest systems we considered ($N=200$, $M=3200$), it takes $\sim10^9$ proposed steps to forget the initial conditions; a typical Monte-Carlo run takes $\sim10^{11}$ proposed steps. The statistical error bars are estimated from several (20--30) independent runs.

Finally, we note that our QMC scheme is much less suitable if the JJ chain does not have ends but is closed into a ring. The details are given in Appendix~\ref{app:ring}.

\section{Detecting the transition}

\subsection{Transition indicator}

Having set up the QMC scheme, one should choose an observable $\hat{O}$, whose average,
\begin{equation}
\langle\hat{O}\rangle
\equiv\frac{\Tr\{\hat{O}\,e^{-\beta{H}_{Q=0}}\}}{\Tr\{e^{-\beta{H}_{Q=0}}\}},
\end{equation}
can distinguish between the superconductor and insulator phases. 
The first excitation energy gap, which shrinks to zero at $N\to\infty$ in the superconductor but remains finite in the insulator, can be calculated from the imaginary-time correlators; however, the gap goes to zero exponentially as the transition is approached from the insulating side, so the transition point cannot be determined precisely.
Charge stiffness, which might seem a natural order parameter of the insulating phase, also turns out to be rather inconvenient (the detailed arguments, which we find quite instructive, are given in Appendix~\ref{app:stiffness}).

We find the most suitable observable to be the total dipole moment of the chain,
\begin{equation}
\hat{d}\equiv\sum_{j=1/2}^{N-1/2}\hat{P}_j,
\end{equation}
whose average is zero, but the zero-temperature fluctuations behave differently in the two phases:
\begin{equation}
\lim_{\beta\to\infty}\langle\hat{d}^2\rangle
%\equiv\frac{\Tr\{\hat{d}^2e^{-\beta{H}_{Q=0}}\}}{\Tr\{e^{-\beta{H}_{Q=0}}\}}
\mathop{\sim}\limits_{N\to\infty}
\left\{\begin{array}{ll}
N^2,&\mbox{superconductor}, \\ N,&\mbox{insulator},
\end{array}\right.
\end{equation}
where the limit $\beta\to\infty$ is taken first.
To see the origin of this scaling, let us first assume to be deep in the superconducting phase. Then in Eq.~(\ref{eq:Htheta=}) one can expand $1-\cos\hat\theta_j\approx\hat\theta_j^2/2$ and evaluate $\langle\hat{d}^2\rangle$ in the harmonic approximation (see Appendix~\ref{app:harmonic}):
\begin{equation}\label{eq:ddsuper=}
\frac{\langle\hat{d}^2\rangle}{(2e)^2}
=\sum_{k=1}^N\frac{1-(-1)^k}2\,\frac{E_J}{(N+1)\omega_k}\,\cot^2\frac{\mu_k}2\coth\frac{\beta\omega_k}2,
\end{equation}
where the normal mode frequency $\omega_k$ and wave vector $\mu_k$ are given by
\begin{equation}
\omega_k=\sqrt{\frac{4(2e)^2E_J\sin^2(\mu_k/2)}{C_g+4C\sin^2(\mu_k/2)}},\quad
\mu_k=\frac{\pi{k}}{N+1}.
\end{equation}
Taking the limit $\beta\to\infty$, we set $\coth(\beta\omega_k/2)\to{1}$. If at large $N$ one replaces the $k$ sum by an integral, it will diverge at the lower limit as $\int{d}\mu/\mu^3$; in fact, the sum is dominated by the first few values of~$k$ and indeed scales as $N^2$. In the insulating phase, the normal modes are gapped, so the frequencies $\omega_k$ saturate to a finite value as $N\to\infty$. This removes one factor of~$N$ from the correlator.

To see how fast the limit $\beta\to\infty$ is reached for a large but finite~$N$, let us go back to Eq.~(\ref{eq:ddsuper=}) with $\coth(\beta\omega_k/2)$ and evaluate the sum focusing on the lowest frequencies:
\begin{subequations}\begin{align}
& \frac{\langle\hat{d}^2\rangle}{(2eN)^2}
=g\,\frac{7\zeta(3)}{2\pi^4}\,
\mathcal{B}\!\left(\frac{N+1}{\beta\sqrt{E_JE_g}}\right),\\
&\mathcal{B}(x)\equiv\frac{8}{7\zeta(3)}
\sum_{m=1}^\infty\frac{1}{(2m-1)^3}
\coth\frac{\pi(2m-1)}{2x},\label{eq:calB=}
\end{align}\end{subequations}
where $\zeta(x)$ is the Riemann $\zeta$~function, and we defined
\begin{equation}\label{eq:KvEg=}
g\equiv\pi\sqrt{E_J/E_g},\quad 
v\equiv\pi\sqrt{E_JE_g},\quad 
E_g\equiv(2e)^2/C_g.
\end{equation}
Here $v$ is the velocity of the low-frequency dispersion $\omega_k\approx{v}\mu_k$ (since the distances are measured in units of the lattice spacing, the velocity has the dimensionality of energy). 
While $\mathcal{B}(0)=1$ strictly at zero temperature, $\mathcal{B}(1/4)=1.00001$, 
$\mathcal{B}(1/2)=1.00356$, and $\mathcal{B}(1)=1.08589$, so in practice the extrapolation $\beta\to\infty$ can be done by taking the temperature an order of magnitude smaller that the first mode frequency $\omega_1\approx\pi\sqrt{E_JE_g}/N$. In the insulating phase, the limit is reached even faster since the lowest excitation energy is finite as $N\to\infty$.

The average $\langle\hat{d}^2\rangle$, being a specific case of the imaginary-time polarization correlator discussed in the end of the previous section, is very well suitable for evaluation by our QMC scheme. 
%Its global character (the sum over~$j$) suppresses statistical Monte-Carlo errors. 
Additional error suppression is achieved by averaging over the imaginary time.

\subsection{Kosterlitz-Thouless scaling}
\label{ssec:RT}

\begin{figure}
\includegraphics[width=0.48\textwidth]{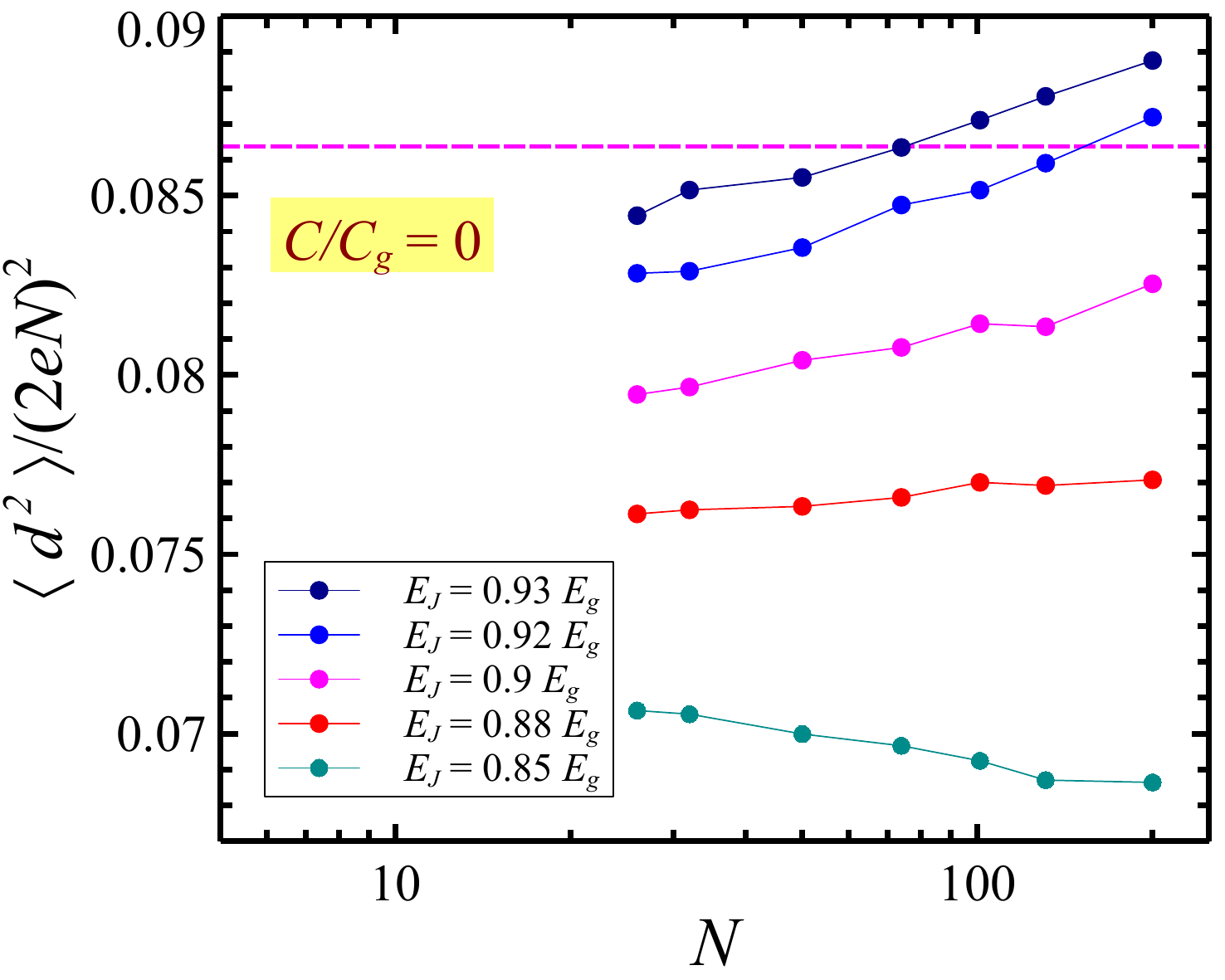}
\caption{\label{fig:correlator}
$N$ dependence of $\langle\hat{d}^2\rangle/(2eN)^2$ for $C/C_g=0$ and different $E_J/E_g$ (from top to bottom): 0.93, 0.92, 0.9, 0.88, 0.85. The data was obtained with $\beta{E}_g=4N$, $\beta{E}_g/M=1/4$. The horizontal dashed line indicates the limiting critical value $7\zeta(3)/\pi^4$.}
\end{figure}

We start with the short-range case $C=0$ and plot Fig.~\ref{fig:correlator} the average $\langle\hat{d}^2\rangle/(2eN)^2$ as a function of $N$ for different~$E_J/E_g$, with the statistical error bars being comparable to the symbol size. The plotted values were obtained for $\beta{E}_g=4N$, $\beta{E}_g/M=1/4$; we checked that increasing $\beta$ or $M$ by a factor of~2 did not change the results, so the limits $\beta\to\infty$ and $M\to\infty$ have been reached. The $N$ dependence in Fig.~\ref{fig:correlator} is very slow, which is typical for the Kosterlitz-Thouless transition. Then, it is helpful to analyze the data using the Kosterlitz-Thouless scaling~\cite{Weber1988, Ceperley1989, Olsson1991, Hsieh2013}.

To establish the scaling of $\langle\hat{d}^2\rangle$, we adopt the low-energy description of the JJ chain in terms of the sine-Gordon model~\cite{Hermon1996, Gurarie2004, Bard2017}. Its Hamiltonian can be written as~\cite{HerbutBook}
\begin{align}
\hat{H}_\mathrm{sG}=v\int{d}x\left[\frac{K}{2\pi}\,\hat\Theta^2
+\frac{\pi}{2K}\left(\frac{1}{2e}\,\frac{\partial\hat{P}}{\partial{x}}\right)^2
\right.\nonumber\\ 
+\left.\frac{y}{a^2}\left(1-\cos\frac{2\pi\hat{P}}{2e}\right)\right].
\label{eq:HsinGordon=}
\end{align}
Here $\hat\Theta=\partial\hat\phi/\partial{x}$ (where $\hat\phi(x)$ is the smoothly varying phase of the superconducting order parameter), $-\partial\hat{P}/\partial{x}$ is the charge density, and $[\hat{P}(x),\hat\Theta(x')]=2ie\delta(x-x')$. Model~(\ref{eq:HsinGordon=}) is ill-defined unless a short-distance regularization is specified. This defines a short-distance cutoff length~$a$. To match the lattice model~(\ref{eq:H=}) the short distance scale should be taken as $a\sim\max\{1,\ell_s\}$; at this scale the parameters $v$ and $K$ of Hamiltonian (\ref{eq:HsinGordon=}) are determined by Eq.~(\ref{eq:KvEg=}) with $K=g$. The parameter~$y$ is known in the limit $C\gg{C}_g$,\begin{equation}\label{eq:y=}
y\sim\sqrt{g\ell_s^3}\,e^{-(8/\pi)g\ell_s},
\end{equation}
the exact value depending on the precise regularization procedure, while for $C\gtrsim{C}_g$ one can only say that $\ln(1/y)\sim{g}$. Eq.~(\ref{eq:y=}) can be understood by choosing a segment $x_0<x<x_0+a$ where the polarization is constant, and treating the rest of the chain at $x<x_0$ and $x>x_0+a$ as external voltage probes with polarization~$P$ (see Appendix~\ref{app:stiffness}). Then, integrating Eq.~(\ref{eq:HsinGordon=}) over $x$ between $x_0$ and $x_0+a$, we obtain the energy $(yv/a)\cos(\pi{P}/e)$ which can be interpreted as the lowest Bloch band dispersion with $P$ playing the role of the quasicharge~\cite{Hermon1996, Gurarie2004}. The instanton calculation of the bandwidth~\cite{Matveev2002, Rastelli2013, Svetogorov2018}, valid for $\ell_s\gg{1}$, gives Eq.~(\ref{eq:y=}).

It is possible to coarse-grain the system by increasing the cutoff $a\to\tilde{a}>a$ and eliminating the modes with high frequencies $v/\tilde{a}<\omega<v/a$. The coarse-grained system is still described by Hamiltonian~(\ref{eq:HsinGordon=}), but with renormalized the coefficients $K$ and~$y$. Their flow with increasing cutoff is governed by the renormalization group (RG) equations~\cite{Kosterlitz1974,HerbutBook}: 
\begin{equation}
\frac{dK}{d\ln{a}}=-\alpha{y}^2,\quad \frac{dy}{d\ln{a}}=(2-K)y.
\end{equation}
Here $\alpha\sim1$ is an unknown numerical factor, whose uncertainty stems from that in the definition of the short-distance cutoff~$a$ [since we have not specified the precise short-range regularization procedure, the scale~$a$ is defined up to a numerical factor, and so is the coefficient $y$ at the cosine term in Eq.~(\ref{eq:HsinGordon=})]. These RG equations should be integrated from $a=a_0\sim\max\{1,\ell_s\}$ with the initial conditions $K=g$ and Eq.~(\ref{eq:y=}), up to $a\sim{N}$ on the superconducting side. On the insulating side, the flow should be stopped at the soliton size determined by the condition $4\pi{K}y\sim1$ (see Appendix~\ref{app:harmonic}). The critical trajectory is given by
\begin{equation}\label{eq:critical}
y(a)=\frac{y(a_0)}{\ln(ea/a_0)},\quad
K-2=\frac{\sqrt{\alpha}\,y(a_0)}{\ln(ea/a_0)}.
\end{equation}

As mentioned in the previous subsection, $\langle\hat{d}^2\rangle$ is determined by a few lowest modes, so on the superconducting side and at the transition itself, it can be found by using Hamiltonian~(\ref{eq:HsinGordon=}) with the cosine expanded to the harmonic order and with renormalized parameters $K,y$ corresponding to the scale $a\sim{N}$.  Performing the standard harmonic calculation 
%fully analogous to the one leading to Eq.~(\ref{eq:ddsuper}) 
(see Appendix~\ref{app:harmonic}), at zero temperature we obtain
\begin{equation}
\frac{\langle\hat{d}^2\rangle}{(2eN)^2}
=\sum_{m=1}^\infty
\frac{4K/\pi^4}{(2m-1)^2\sqrt{(2m-1)^2+y(a\sim{N})}}.
\end{equation}
On the critical trajectory~(\ref{eq:critical}), $y(a\sim{N})\propto1/\ln{N}$ while $K$ flows to~$2$, so  $\langle\hat{d}^2\rangle/(2eN)^2$ attains a universal value $7\zeta(3)/\pi^4=0.086\ldots$. Therefore,
\renewcommand{\labelenumi}{(\roman{enumi})}
\begin{enumerate}
\item
in the superconducting phase, $\langle\hat{d}^2\rangle/(2eN)^2$ monotonously increases with $N$ to some limiting value exceeding $7\zeta(3)/\pi^4$;
\item
in the insulating phase, the flow turns downwards at some value of $N$ which is exponentially large in the distance to the critical point, and the value of $\langle\hat{d}^2\rangle/(2eN)^2$ at the downturn is smaller than $7\zeta(3)/\pi^4$;
\item
it is impossible for $\langle\hat{d}^2\rangle/(2eN)^2$ to exceed $7\zeta(3)/\pi^4$ and subsequently turn downwards.
\end{enumerate}
In Fig.~\ref{fig:correlator}, the curves for $E_J/E_g=0.92$ and 0.88 fall into cases (i) and~(ii), respectively. The curve for $E_J/E_g=0.90$ is uncertain, and larger~$N$ is needed to draw a definite conclusion. This determines the error bars of our procedure. As a result, we obtain the critical value for $C=0$, $g_c=2.98\pm0.03$. This value is fully consistent with $2.97\pm0.03$ of Ref.~\cite{Roscilde2016}. Ref.~\cite{Danshita2011} gives $g_c=3.024$ with the statistical error of $\pm0.004$; at the same time, a systematic error of about 3\% favoring the insulating phase, was discussed in that paper, making the result also consistent with ours. These values are incompatible with $2.50\pm0.08$, where a minimum of the ground state fidelity was observed in Ref.~\cite{Pino2016}.

\subsection{Behavior at large~$\ell_s$}
\label{sec:asymptote}

\begin{figure}
\includegraphics[width=0.45\textwidth]{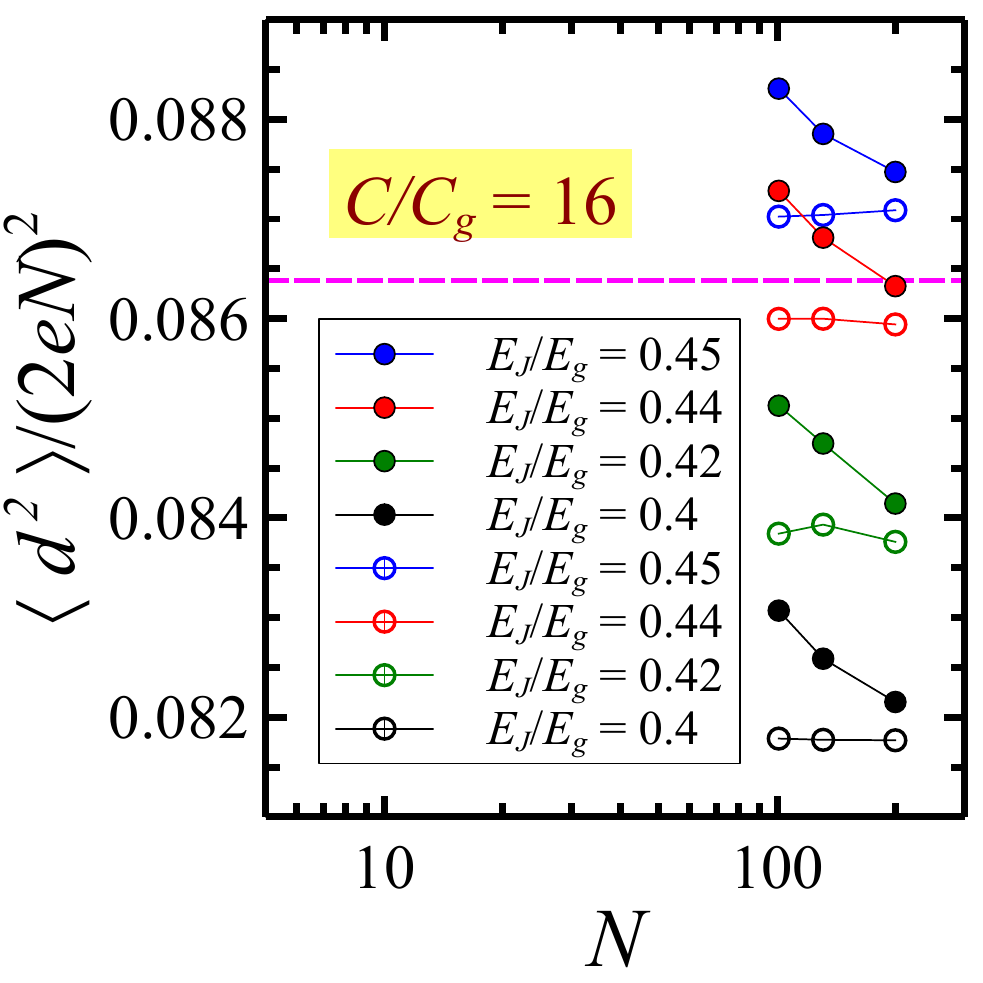}
\caption{\label{fig:largels}
Filled circles: $N$ dependence of $\langle\hat{d}^2\rangle/(2eN)^2$ for $C/C_g=16$ ($\ell_s=4$) and different $E_J/E_g$ (from top to bottom): 0.45, 0.44, 0.42, 0.4. The data was obtained with $\beta{E}_g=4N$, $\beta{E}_g/M=1/4$. Open circles: the same data after subtraction of $\delta_N$ (see text for details). The horizontal dashed line indicates the limiting critical value $7\zeta(3)/\pi^4$.}
\end{figure}

Upon increasing~$\ell_s$, one needs larger and larger sizes to resolve the asymptotic behavior at $N\to\infty$. In Fig.~\ref{fig:largels} we show the $N$~dependence of $\langle\hat{d}^2\rangle/(2eN)^2$ for a few values of $E_J/E_g$ at $\ell_s=4$ (filled circles). This dependence shows a contribution on top of the slow Kosterlitz-Thouless scaling, which prevents us from applying directly the method discussed in the preceding subsection. This contribution appears to be non-critical, so its origin can be understood using the superconducting expression~(\ref{eq:ddsuper=}): at finite $C$, it contains a subleading correction $\sim(\ell_s/N)^2\ln(N/\ell_s)$. Let us denote by $\delta_N$ the difference between expression~(\ref{eq:ddsuper=}) at finite~$N$ and its limit at $N\to\infty$ for $E_J=(2/\pi)^2E_g$ (giving $g=2$). Assuming different contributions to scaling to be additive near the critical fixed point, we simply subtract $\delta_N$ from the data. The result is shown in Fig.~\ref{fig:largels} by the open circles. The corrected data is rather flat, so we determine the critical value of $E_J$ as the one giving $\langle\hat{d}^2\rangle/(2eN)^2=7\zeta(3)/\pi^4$ (in practice, we interpolate from the four sets shown in Fig.~\ref{fig:largels}). This gives $E_J/E_g=0.444$ for $\ell_s=4$. Other points in Fig.~\ref{fig:JJchain} with $\ell_s>1$ are also obtained using this procedure.

At $\ell_s\to\infty$, the critical value $g_c\to2$. How is this asymptote approached? Looking at Eqs.~(\ref{eq:y=}) and~(\ref{eq:critical}), one could think that this approach is exponential in~$\ell_s$~\cite{Korshunov1989}. However, the $\ell_s$ dependence in Fig.~\ref{fig:JJchain} is clearly slower than exponential. This points at another contribution to renormalization of~$K$, not accounted for by the Kosterlitz-Thouless RG where $K$~is renormalized by bound vortex-antivortex pairs.

If one goes beyond the harmonic approximation in Hamiltonian~(\ref{eq:Htheta=}) and expands the Josephson term to the next order, $1-\cos\hat\theta_j\approx\hat\theta_j^2/2-\hat\theta_j^4/24$, the harmonic mode frequencies are shifted by the Kerr effect~\cite{Weissl2015, Krupko2018}. At zero temperature, average of $-\hat\theta_j^4/24$ over the zero-point oscillations for a single junction produces an effective correction to the Josephson energy, $\delta{E}_J=-(1/4)\sqrt{E_J(2e)^2/C}$, which can be translated into a correction to the initial condition for~$K$: instead of $K=g$, it should be $K=g-\pi/(8\ell_s)$. The transition occurs when the renormalized~$K$ is equal to~2, which gives the critical value
\begin{equation}\label{eq:largels=}
g_c=2+\frac\pi{8\ell_s}+O(1/\ell_s^2).
\end{equation}
This expression is plotted in Fig.~\ref{fig:JJchain} by the dashed line and matches remarkably well the QMC result down to $\ell_s\approx{1}$. We emphasize that this Kerr renormalization is a short-distance effect and is not captured by the Kosterlitz-Thouless RG.

\section{Conclusions}

We have developed a novel imaginary path-integral QMC scheme in the charge representation which can efficiently handle quantum phase models with arbitrary electrostatic interactions. We applied this method to the superconductor-insulator transition in  a dissipationless and disorder-free Josephson junction chain characterized by two capacitances, where the Coulomb interaction between the charges decays exponentially with distance. We have benchmarked our method with the known results for the special case of contact interaction, when the chain is equivalent to the Bose-Hubbard model at large integer filling. At screening lengths $\ell_s\gtrsim1$, the transition line is governed by short-distance renormalizations due to the weak Kerr nonlinearity of each junction, not captured by the Kosterlitz-Thouless renormalization group.

\acknowledgements

The authors acknowledge illuminating discussions with
F.~Alet,
D.~A.~Ivanov,
G.~Rastelli,
T.~Roscilde,
A.~Shnirman,
A.~E.~Svetogorov,
and M.~E.~Zhitomirsky.
This work is supported by the project THERMOLOC (ANR-16-CE30-0023-02) of the French National Research Agency (ANR). 
P. A. acknowledges support by the H2020 European programme under the project TWINFUSYON (GA692034).
Most of the computations were performed using the Froggy platform of the CIMENT infrastructure (https://ciment.ujf-grenoble.fr), which is supported by the Rh\^one-Alpes region (grant CPER07-13 CIRA) and the project Equip@Meso (ANR-10-EQPX-29-01) of the ANR.

\appendix

\section{Josephson junction ring}
\label{app:ring}

Closing the chain into a ring corresponds to adding a junction between islands $n=N$ and $n=0$. This introduces no new degrees of freedom, but (i)~it modifies four elements of the capacitance matrix, $C_{00}=C_{NN}=C_g-2C$, $C_{0N}=C_{N0}=-C$, and (ii)~it introduces an extra term in the Josephson part of the Hamiltonian, $E_J[1-\cos(\phi_N-\phi_0-\varphi)]$, if the ring is pierced by a magnetic flux $\varphi$ (in the units of flux quantum divided by $2\pi$). In the variables~$\theta_j$ which are defined in the same way as for the open chain, the Josephson part of the Hamiltonian becomes
\begin{align}
\hat{H}_J={}&{}\sum_{j=1/2}^{N-1/2}E_J(1-\cos\theta_j)\nonumber\\
&{}+E_J[1-\cos(\theta_{1/2}+\ldots+\theta_{N-1/2}-\varphi)].
\end{align}
When constructing the path integral, one can no longer evaluate the matrix element of $e^{-\varepsilon\hat{H}_J}$ at different junctions independently. Still, it can be calculated by introducing additional decoupling variables:
\begin{align}
&\int\limits_0^{2\pi}\prod_{j=1/2}^{N-1/2}\frac{d\theta_j}{2\pi}\,
e^{i\sum_j(l_j-l_j')\theta_j}\times{}\nonumber\\
&\qquad{}\times\exp\left[\varepsilon{E}_J\sum_j\cos\theta_j
+\varepsilon{E}_J\cos\left(\sum_j\theta_j-\varphi\right)\right]\nonumber\\
&=\sum_{k=-\infty}^\infty\int\limits_0^{2\pi}\frac{d\vartheta}{2\pi}
\prod_{j=1/2}^{N-1/2}\frac{d\theta_j}{2\pi}\,
e^{i\sum_j(l_j-l_j')\theta_j}\times{}\nonumber\\
&\qquad\qquad{}\times
e^{ik(\theta_{1/2}+\ldots+\theta_{N-1/2}-\varphi-\vartheta)}\times{}\nonumber\\
&\qquad\qquad{}\times\exp\left[\varepsilon{E}_J\sum_j\cos\theta_j
+\varepsilon{E}_J\cos\vartheta\right]\nonumber\\
&{}=\sum_{k=-\infty}^\infty e^{ik\varphi}\,I_k(\varepsilon{E}_J)
\prod_jI_{l_j-l_j'-k}(\varepsilon{E}_J).
\end{align}
Then, instead of Eqs.~(\ref{eq:QMC}) we have
\begin{subequations}\begin{align}
&\Tr\left\{e^{-\beta\hat{H}}\right\}=e^{-\beta(N+1)E_J}\times{}\nonumber\\
&\hspace*{2cm}{}\times\lim_{M\to\infty}\sum_{\{l_{jm},k_m\}=-\infty}^\infty
W_CW_J'e^{i\varphi\sum_mk_m},\\
&W_J'=\prod_{m=0}^{M-1}I_{k_m}(\beta{E}_J/M)
\prod_{j=1/2}^{N-1/2}I_{l_{jm}-l_{j,m+1}-k_m}(\beta{E}_J/M).
\end{align}\end{subequations}
This expression, formally as good as Eqs.~(\ref{eq:QMC}), is much less convenient from the practical point of view. First, the summand is no longer positive due to the factors $e^{ik_m\varphi}$, which leads to strong cancellations (sign problem). Second, we do not see an efficient way to sample configurations: since the variables $k_m$ appear in many Bessel functions, even a small modification of the configuration may lead to a strong modification of the weight resulting in a low acceptance probability.

\section{Charge stiffness}
\label{app:stiffness}

To define the charge stiffness, one should choose two arbitrary islands $n_1,n_2$ and modify the Coulomb part of Hamiltonian~(\ref{eq:H=}) as
\begin{align}
\hat{H}_\kappa={}&{}\sum_{n,n'=0}^N\frac{C_{nn'}^{-1}}2\,(\hat{q}_n-\kappa_n)
(\hat{q}_{n'}-\kappa_{n'})\nonumber\\
{}&{}+\sum_{n=1}^NE_J[1-\cos(\hat\phi_n-\hat\phi_{n-1})],
\label{eq:Hkappa=}
\end{align}
with $\kappa_n=\kappa\delta_{nn_1}-\kappa\delta_{nn_2}$. The ground state energy in the $Q=0$ sector, $E_0(\kappa)$, is a periodic function of the offset charge~$\kappa$ with period $2e$ since $\hat{q}_{n_1},\hat{q}_{n_2}$ can be shifted by $\pm2e$ still conserving the total charge. The charge stiffness is defined as
\begin{equation}\label{eq:Kdef=}
\mathcal{K}_{n_1n_2}=
\left.-\frac{\partial^2{E}_0}{\partial\kappa^2}\right|_{\kappa=0},
\end{equation}
and can be viewed as the inverse capacitance of the system between the two points $n_1,n_2$. 
Using perturbation theory in $\kappa$, definition~(\ref{eq:Kdef=}) can be identically rewritten in the form of an imaginary-time correlator, perfectly suitable for calculation in our QMC scheme:
\begin{align}
\mathcal{K}_{n_1n_2}={}&{}
C_{n_1n_1}^{-1}+C_{n_2n_2}^{-1}-C_{n_1n_2}^{-1}-C_{n_2n_1}^{-1}
\nonumber\\
&{}-\lim_{\beta\to\infty}\int_0^\beta d\tau
\left\langle{e}^{\tau\hat{H}_{Q=0}}\hat{V}_{n_1n_2}
{e}^{-\tau\hat{H}_{Q=0}}\hat{V}_{n_1n_2}\right\rangle.\label{eq:Kcorr=}
%&{}-\lim_{\beta\to\infty}\int_0^\beta d\tau
%\frac{\Tr\{e^{(\tau-\beta)\hat{H}_{Q=0}}\hat{V}_{n_1n_2}
%{e}^{-\tau\hat{H}_{Q=)}}\hat{V}_{n_1n_2}\}}%
%{\Tr\{e^{-\beta\hat{H}_{Q=0}}\}}.\label{eq:Kcorr=}
\end{align}
where
\begin{equation}
\hat{V}_{n_1n_2}=\left.\frac{\partial\hat{H}}{\partial\kappa}\right|_{\kappa=0}
=\sum_{n'=0}^N\left(C_{n_1n'}^{-1}-C_{n_2n'}^{-1}\right)\hat{q}_{n'}
\end{equation}
is nothing but the voltage between the islands $n_1$ and~$n_2$. Thus, it is helpful to think of these two sites as attached to voltage probes.
Naively, one might expect that at $N\to\infty$ the charge stiffness should be finite in the insulating phase and vanish in the superconducting phase.

\begin{figure}
\includegraphics[width=0.48\textwidth]{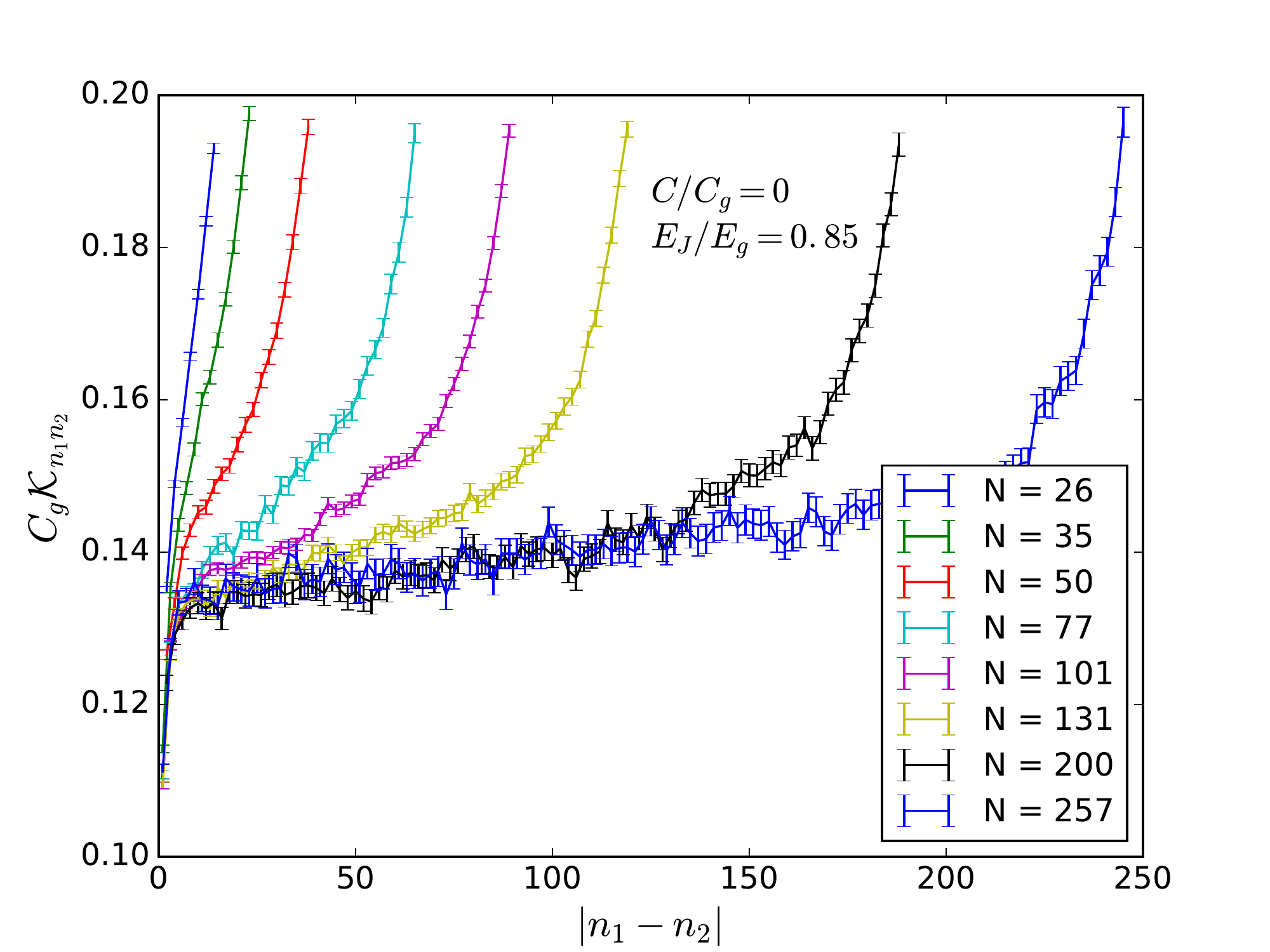}\
\includegraphics[width=0.48\textwidth]{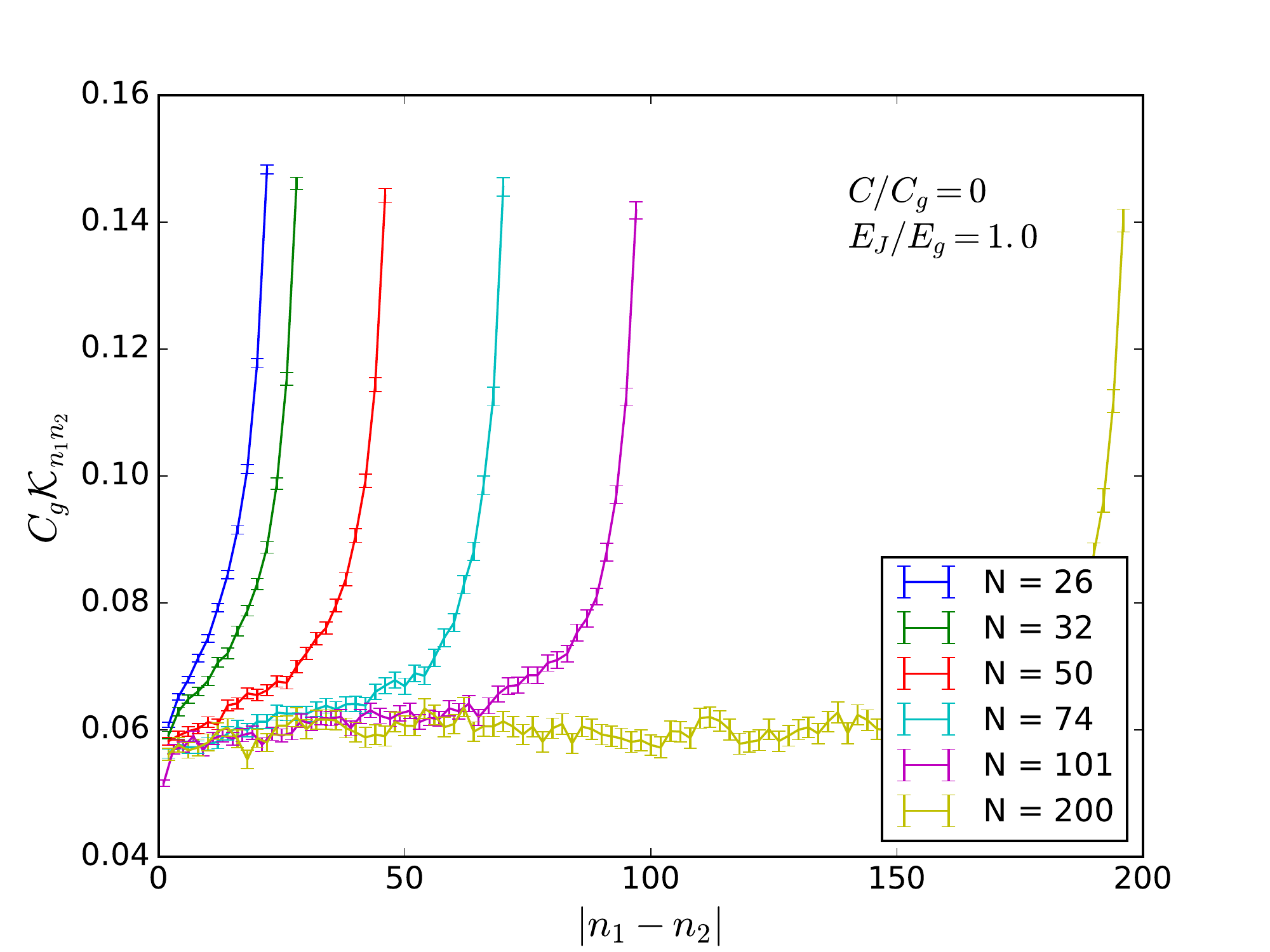}
\caption{\label{fig:stiffness}
Dimensionless charge stiffness $C_g\mathcal{K}_{n_1n_2}$ with $n_2=N-n_1$ for two values of $E_J/E_g=0.85,\,1.0$ (upper and lower panels, respectively) and different chain lengths~$N$ at $C=0$.
}
\end{figure}

Let us take $C=0$ and two values of $E_J/E_g=0.85,\,1.0$, corresponding to the insulating and superconucting phases, respectively. Fixing $n_2=N-n_1$, we show the calculated charge stiffness $\mathcal{K}_{n_1n_2}$ in the natural units of $1/C_g$ and different chain lengths~$N$ in Fig.~\ref{fig:stiffness}. First, we observe that the stiffness remains finite and relatively large when the voltage probes are attached to the ends of the chain, $n_1\to{0}$, $n_2\to{N}$. More puzzling, even when the voltage probes are placed in the bulk of the chain, $\mathcal{K}_{n_1n_2}$ tends to a small but finite value as $N\to\infty$.

To clarify these results, let us recall that the ground state energy $E_0(\kappa)$ can also be viewed as the dispersion of lowest Bloch band. Indeed, if external wires are attached to the islands $n_1$ and $n_2$, the corresponding phases $\phi_{n_1},\phi_{n_2}$ become non-compact (that is, all values on the whole real axis $\mathbb{R}$ become physically distinct). In other words, the $\{\phi_n\}$ space instead of a $(N+1)$-dimensional torus $\mathbb{T}_{N+1}$ becomes $\mathbb{R}^2\otimes\mathbb{T}_{N-1}$. When passing from $(\phi_0,\ldots,\phi_N)$ to $(\theta_{1/2},\ldots,\theta_{N-1/2},\Phi)$, the overall phase $\Phi$ becomes non-compact, conjugate to the continuous total charge~$Q$ which is still conserved since the Hamiltonian does not depend on~$\Phi$. The remaining $(\theta_{1/2},\ldots,\theta_{N-1/2})$ lie on an $N$-dimentional cylinder $\mathbb{R}\otimes\mathbb{T}_{N-1}$, with the non-compact direction corresponding to
\[
\phi_{n_2}-\phi_{n_1}=\sum_{j=n_1+1/2}^{n_2-1/2}\theta_j.
\]
The Josephson energy is still a periodic function of all~$\theta_j$, so along this non-compact direction the Hamiltonian has a discrete translation symmetry. Then $\kappa$ can be viewed as the quasicharge quantum number arising by virtue of the Bloch theorem, while Hamiltonian~(\ref{eq:Hkappa=}) is precisely the Hamiltonian for the periodic part of the Bloch function. The charge stiffness is just the band curvature at the bottom, and Eq.~(\ref{eq:Kcorr=}) is the analog of the $\mathbf{k}\cdot\mathbf{p}$ perturbation theory, a common tool in the band theory of solids. If the voltage probes are viewed as one-dimensional wires in which a polarization~$P$ is created, the offset charges $\kappa_{n_1,n_2}=\pm\kappa$ entering Hamiltonian~(\ref{eq:Hkappa=}) can be associated with the boundary charges of the polarized wires, $\kappa=P$.

For a finite-length chain with $E_J$~sufficiently large, the phase is almost classical, the lowest Bloch band is sinusoidal, and its small bandwidth is determined by tunneling between two neighboring minima of the Josephson energy. For example, one can consider the minimum with all $\theta_j=0$, and the neighboring one with $\theta_j=2\pi\delta_{jj_0}$, $n_1<j_0<n_2$ (note that $n_2-n_1$ possible values of $j_0$ correspond to a single point on the $\mathbb{R}\otimes\mathbb{T}_{N-1}$ cylinder). Tunneling between neighboring minima is called a quantum phase slip, and the Bloch bandwidth can be calculated using the instanton approach~\cite{Hekking1997, Matveev2002, Buchler2004, Rastelli2013, Svetogorov2018}.
The bandwidth corresponds to the amplitude of a quantum phase slip at any junction between the voltage probes $n_1,n_2$. Equivalently, it is given by the density of vortices of the classical $XY$ model in the imaginary-time direction, whose spatial position is between the voltage probes.
Naively, one would expect $\mathcal{K}_{n_1n_2}$ to be finite in the insulating phase (characterized by a finite density of unpaired vortices) and to decay as $\mathcal{K}_{n_1n_2}\propto|n_1-n_2|/N^K$ in the superconducting phase (where each vortex has a large self-energy $\propto1/N^K$).

First, in Fig.~\ref{fig:stiffness} we observe that the stiffness remains finite and relatively large when the voltage probes are attached to the ends of the chain, $n_1\to{0}$, $n_2\to{N}$. Ths happens because the action of a phase slip occurring near one of the chain ends is not proportional to $\ln{N}$, but is cut off by the distance to the end. This effect was discussed in Ref.~\cite{Buchler2004} for a superconducting wire. Thus, there is a finite density of unpaired vortices near the chain ends even in the superconducting phase.

Second, even when the probes are placed in the bulk of the chain, $\mathcal{K}_{n_1n_2}$ tends to a small but finite value as $N\to\infty$. This happens because in addition to the single-vortex contribution to the phase slip amplitude, there is another contribution due to bound vortex-antivortex pairs where the vortex resides on one side of a voltage probe and the antivortex on the on the other side, and thus the phase $2\pi$ is accumulated on the probe as $\tau$ goes through the pair. This pair contribution to the amplitude is subleading in the vortex fugacity and thus quickly decreases with increasing $E_J$, but it does not scale with $N$, nor with the probe separation $|n_1-n_2|$ (except for small $|n_1-n_2|\sim{1}$, which corresponds to the size of the bound vortex-antivortex pair).

\begin{figure}
\includegraphics[width=0.48\textwidth]{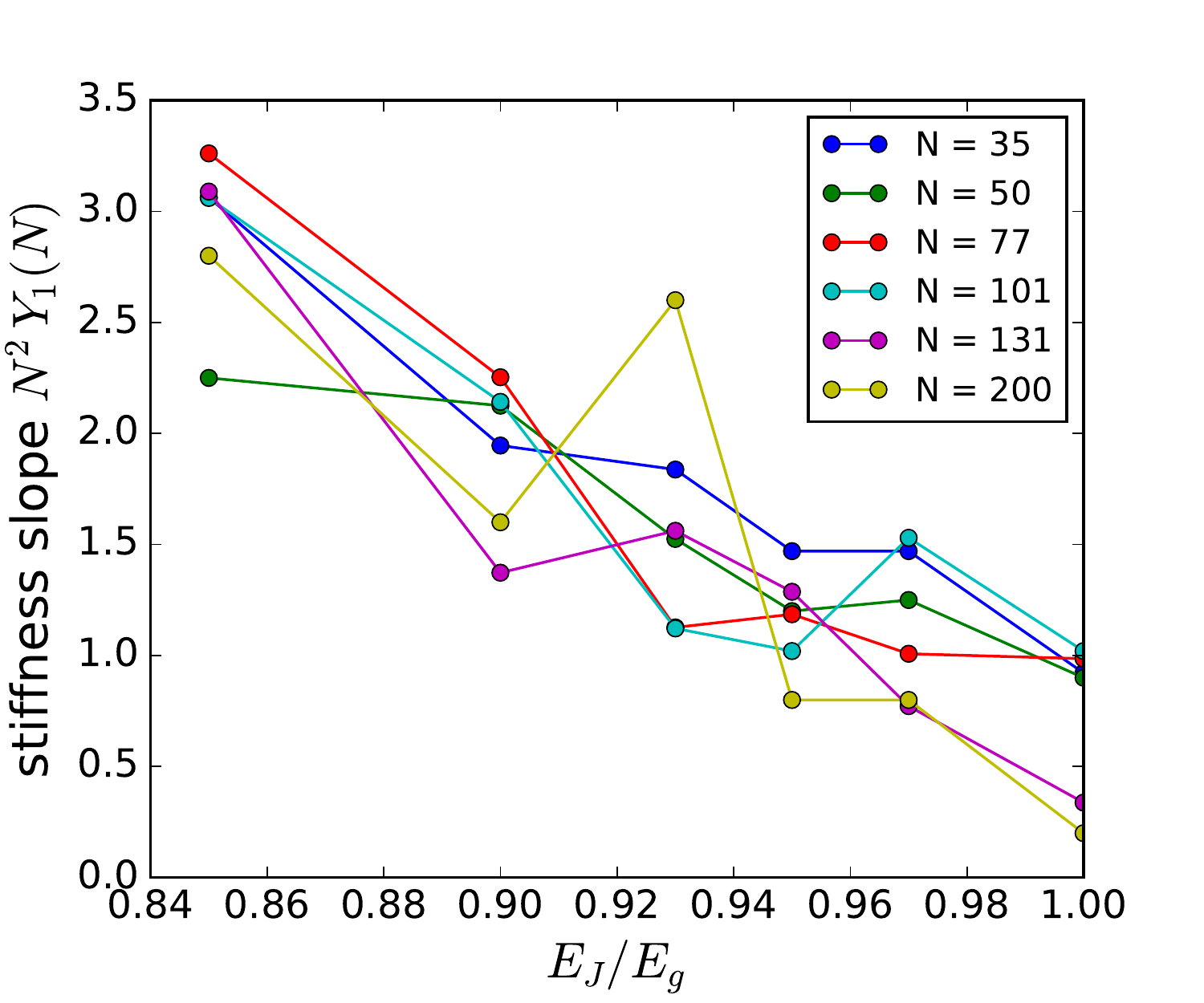}
\caption{\label{fig:stiffnessslope}
The rescaled slope $N^2Y_1(N)$ of the charge stiffness [Eq.~(\ref{eq:stiffness_slope})] as a function of $E_J/E_g$ for different~$N$.}
\end{figure}

Thus, for a given $N$, one can fit $\mathcal{K}_{n_1n_2}$ for $|n_1-n_2|$ sufficiently far from~$N$ as a function of $|n_1-n_2|$ by a linear function,
\begin{equation}\label{eq:stiffness_slope}
\mathcal{K}_{n_1n_2}=Y_0+|n_1-n_2|\,Y_1,
\end{equation}
and try to extract $Y_1(N)\propto{1}/N^K$ in the superconducting phase, and $Y_1(N)\propto{1}/(N^2\ln{N})$ at the transition. However, the uncertainty of thus obtained exponent turns out to be too high. We plot $N^2Y_1(N)$ as a function of $E_J$ for different~$N$ in Fig.~\ref{fig:stiffnessslope}. Ideally, one would hope to see a family of smooth curves, the steeper the larger is $N$, all crossing at one point, the critical value of~$E_J$. But the data is too noisy to be useful in practice.

\section{Harmonic calculation}\label{app:harmonic}

To handle the harmonic part of Hamiltonian~(\ref{eq:Htheta=}),
\begin{equation}
\hat{H}_{Q=0}^{(2)}=\sum_{j,j'=1/2}^{N-1/2}\frac{D_{jj'}}2\,\hat{P}_j\hat{P}_{j'}
+\frac{E_J}2\sum_{j=1/2}^{N-1/2}\hat\theta_j^2,
\end{equation}
we need to diagonalize the dipole-dipole matrix~(\ref{eq:Djj=}). Let us start from the tridiagonal capacitance matrix, which can be written in terms of its eigenvectors $u_{nk}$ and eigenvalues $C_k$ as
\begin{subequations}\begin{align}
&C_{nn'}=\sum_{k=0}^NC_ku_{nk}u_{n'k},\quad C_k=C_g+4C\sin^2\frac{\mu_k}2,\\
&u_{nk}=\sqrt{\frac{2-\delta_{k0}}{N+1}}\cos[\mu_k(n+1/2)],\quad
\mu_k\equiv\frac{\pi{k}}{N+1}.
\end{align}\end{subequations}
Inverting $C_{nn'}$ and using the definition~(\ref{eq:Djj=}), we straightforwardly obtain
\begin{subequations}\begin{align}
&D_{jj'}=\sum_{k=1}^ND_k\bar{u}_{jk}\bar{u}_{j'k},\quad
D_k=\frac{4\sin^2(\mu_k/2)}{C_g+4C\sin^2(\mu_k/2)},\\
&\bar{u}_{nk}=\sqrt{\frac{2}{N+1}}\sin[\mu_k(j+1/2)].
\end{align}\end{subequations}
This gives the harmonic Hamiltonian in terms of the normal mode creation and annihilation operators $\hat{b}_k^\dagger, \hat{b}_k$:
\begin{subequations}\begin{align}
&\hat{H}_{Q=0}^{(2)}=\sum_{k=1}^N\omega_k
\left(\hat{b}_k^\dagger\hat{b}_k+\frac{1}2\right),\quad
\omega_k=\sqrt{(2e)^2D_kE_J},\\
&\hat\theta_j=\sum_k\left[\frac{(2e)^2D_k}{4E_J}\right]^{1/4}\bar{u}_{jk}
\left(\hat{b}_k+\hat{b}_k^\dagger\right),\\
&\hat{P}_j=2ie\sum_k\left[\frac{E_J}{4(2e)^2D_k}\right]^{1/4}\bar{u}_{jk}
\left(\hat{b}_k-\hat{b}_k^\dagger\right).
\end{align}\end{subequations}
Noting that
\[
\sum_{j=1/2}^{N-1/2}\bar{u}_{jk}=\sqrt{\frac{2}{N+1}}\,\frac{1-(-1)^k}2\cot\frac{\mu_k}2.
\]
we arrive at Eq.~(\ref{eq:ddsuper=}) by a straightforward calculation.

For Hamiltonian~(\ref{eq:HsinGordon=}) with the cosine expanded to quadratic order, we have
\begin{subequations}\begin{align}
&\hat{H}_\mathrm{sG}^{(2)}=\sum_{k=1}^N\tilde\omega_k
\left(\hat{b}_k^\dagger\hat{b}_k+\frac{1}2\right),\quad
\tilde\omega_k=\sqrt{v^2\mu_k^2+4\pi{K}y\,\frac{v^2}{a^2}},\\
&\hat\Theta(x)=\sum_k\sqrt{\frac{\pi\tilde\omega_k}{(N+1)vK}}
\left(\hat{b}_k+\hat{b}_k^\dagger\right)\sin\mu_kx,\\
&\hat{P}(x)=2ie\sum_k\sqrt{\frac{vK}{(N+1)\pi\tilde\omega_k}}
\left(\hat{b}_k-\hat{b}_k^\dagger\right)\sin\mu_kx.
\end{align}\end{subequations}
Here we used the zero-current boundary conditions, $\hat{P}(x=0)=\hat{P}(x=N+1)=0$. The gap in $\tilde\omega_k$ determines the soliton size $a/\sqrt{4\pi{K}y}$. Evaluation of $\langle\hat{d}^2\rangle$ gives
\begin{equation}
\frac{\langle\hat{d}^2\rangle}{(2e)^2}=\sum_{k=1}^\infty
\frac{1-(-1)^k}2\,\frac{4vK}{(N+1)\pi\tilde\omega_k\mu_k^2}
\coth\frac{\beta\tilde\omega_k}2.
\end{equation}

\end{document}